\documentclass[]{aastex701}

\usepackage{booktabs}

\begin{document}

\title{Delayed, Line-dependent Reorganization of Pore Oscillations Following a Compact C8.2 Flare Observed by {\sc Sunrise iii}/SUSI}

\author[orcid=0000-0002-7711-5397]{Shahin Jafarzadeh}
\affiliation{Astrophysics Research Centre, School of Mathematics and Physics, Queen’s University Belfast, Belfast, BT7 1NN, UK}
\affiliation{Max Planck Institute for Solar System Research, Justus-von-Liebig-Weg 3, 37077 G\"{o}ttingen, Germany}
\email[show]{shahin.jafarzadeh@qub.ac.uk}  

\author[orcid=0000-0002-9155-8039]{David B. Jess} 
\affiliation{Astrophysics Research Centre, School of Mathematics and Physics, Queen’s University Belfast, Belfast, BT7 1NN, UK}
\affiliation{Department of Physics and Astronomy, California State University Northridge, Northridge, CA 91330, USA}
\email{d.jess@qub.ac.uk}  

\author[orcid=0000-0002-5365-7546]{Marco Stangalini} 
\affiliation{ASI Italian Space Agency, Via del Politecnico snc, I-00133 Rome, Italy}
\email{marco.stangalini@asi.it}   

\author[orcid=0000-0001-9315-7899]{Lyndsay Fletcher} 
\affiliation{Scottish Universities Physics Alliance, School of Physics and Astronomy, University of Glasgow, Glasgow, UK}
\affiliation{Rosseland Centre for Solar Physics, University of Oslo, P.O.Box 1029 Blindern, NO-0315 Oslo, Norway}
\email{lyndsay.fletcher@glasgow.ac.uk}  

\author[orcid=0000-0001-5031-1892]{Ryan O. Milligan} 
\affiliation{Astrophysics Research Centre, School of Mathematics and Physics, Queen’s University Belfast, Belfast, BT7 1NN, UK}
\email{R.Milligan@qub.ac.uk}  

\author[orcid=0000-0002-4819-1884]{Paulo J. A. Sim{\~{o}}es} 
\affiliation{Centro de R{\'{a}}dio Astronomia e Astrof{\'{i}}sica Mackenzie, Escola de Engenharia, Universidade Presbiteriana Mackenzie, 01302-907 S{\~{a}}o Paulo, Brazil}
\affiliation{Scottish Universities Physics Alliance, School of Physics and Astronomy, University of Glasgow, Glasgow, UK}
\email{paulo@craam.mackenzie.br}

\author[orcid=0000-0001-8556-470X]{Peter H. Keys}
\affiliation{Astrophysics Research Centre, School of Mathematics and Physics, Queen’s University Belfast, Belfast, BT7 1NN, UK}
\email{p.keys@qub.ac.uk}  

\author[orcid=0000-0002-3418-8449,sname='Solanki']{Sami~K.~Solanki} \affiliation{Max Planck Institute for Solar System Research, Justus-von-Liebig-Weg 3, 37077 G\"{o}ttingen, Germany}\email{solanki@mps.mpg.de}	

\author[orcid=0000-0003-3490-6532,sname='Smitha']{H.~N.~Smitha} \affiliation{Max Planck Institute for Solar System Research, Justus-von-Liebig-Weg 3, 37077 G\"{o}ttingen, Germany}\email{narayanamurthy@mps.mpg.de}	

\author[orcid=0000-0003-1459-7074,sname='Lagg']{Andreas~Lagg} \affiliation{Max Planck Institute for Solar System Research, Justus-von-Liebig-Weg 3, 37077 G\"{o}ttingen, Germany}\email{lagg@mps.mpg.de}	

\author[orcid=0000-0002-9972-9840,sname='Gandorfer']{Achim~Gandorfer} \affiliation{Max Planck Institute for Solar System Research, Justus-von-Liebig-Weg 3, 37077 G\"{o}ttingen, Germany}\email{gandorfer@mps.mpg.de}

\author[orcid=0009-0009-4425-599X,sname='Feller']{Alex~Feller} \affiliation{Max Planck Institute for Solar System Research, Justus-von-Liebig-Weg 3, 37077 G\"{o}ttingen, Germany}\email{feller@mps.mpg.de}	

\author[orcid=0000-0003-1409-1145,sname='Iglesias']{Francisco~A.~Iglesias} \affiliation{Max Planck Institute for Solar System Research, Justus-von-Liebig-Weg 3, 37077 G\"{o}ttingen, Germany}\affiliation{Grupo de Estudios en Heliofísica de Mendoza, CONICET, Universidad de Mendoza, Boulogne sur Mer 683, 5500 Mendoza, Argentina}\email{iglesias@mps.mpg.de}	

\author[orcid=0000-0001-6317-4380,sname='Riethmüller']{Tino~L.~Riethmüller} \affiliation{Max Planck Institute for Solar System Research, Justus-von-Liebig-Weg 3, 37077 G\"{o}ttingen, Germany}\email{riethmueller@mps.mpg.de}	

\author[sname='Grauf']{Bianca~Grauf} \affiliation{Max Planck Institute for Solar System Research, Justus-von-Liebig-Weg 3, 37077 G\"{o}ttingen, Germany}\email{grauf@mps.mpg.de}	

\author[orcid=0000-0001-6029-7529,sname='Hoelken']{Johannes~Hoelken} \affiliation{Max Planck Institute for Solar System Research, Justus-von-Liebig-Weg 3, 37077 G\"{o}ttingen, Germany}\email{hoelken@mps.mpg.de}
\author[orcid=0000-0002-3387-026X,sname='del~Toro~Iniesta']{Jose~Carlos~del~Toro~Iniesta} \affiliation{Instituto de Astrofísica de Andalucía, CSIC, Glorieta de la Astronomía s/n, 18008 Granada, Spain}\affiliation{Spanish Space Solar Physics Consortium}\email{jti@iaa.es}
\author[orcid=0000-0002-5054-8782,sname='Katsukawa']{Yukio~Katsukawa} \affiliation{National Astronomical Observatory of Japan, 2-21-1 Osawa, Mitaka, Tokyo 181-8588, Japan}\affiliation{Department of Astronomy, The University of Tokyo, 7-3-1, Hongo, Bunkyo-ku, Tokyo 113-0033, Japan}\affiliation{Department of Astronomical Science, The Graduate University for Advanced Studies (SOKENDAI), 2-21-1 Osawa, Mitaka, Tokyo 1818588, Japan}\email{yukio.katsukawa@nao.ac.jp}	
\author[orcid=0000-0002-0787-8954,sname='Bernasconi']{Pietro~Bernasconi} \affiliation{Johns Hopkins University Applied Physics Laboratory, 11100 Johns Hopkins Road, Laurel, Maryland, USA}\email{pietro.bernasconi@jhuapl.edu}	
\author[sname='Berkefeld']{Thomas~Berkefeld} \affiliation{Institut für Sonnenphysik (KIS), Georges-Köhler-Allee 401a, 79110 Freiburg, Germany}\email{thomas.berkefeld@leibniz-kis.de}	

\author[orcid=0000-0001-9228-3412,sname='Álvarez-Herrero']{Alberto~Álvarez-Herrero} \affiliation{Instituto Nacional de T\'ecnica Aeroespacial (INTA), Ctra. de Ajalvir, km. 4, E-28850 Torrejón de Ardoz, Spain}\affiliation{Spanish Space Solar Physics Consortium}\email{alvareza@inta.es}	
\author[orcid=0000-0001-5616-2808,sname='Kubo']{Masahito~Kubo} \affiliation{National Astronomical Observatory of Japan, 2-21-1 Osawa, Mitaka, Tokyo 181-8588, Japan}\email{masahito.kubo@nao.ac.jp}	
\author[orcid=0000-0001-8829-1938,sname='Orozco~Suárez']{David~Orozco~Suárez} \affiliation{Instituto de Astrofísica de Andalucía, CSIC, Glorieta de la Astronomía s/n, 18008 Granada, Spain}\affiliation{Spanish Space Solar Physics Consortium}\email{orozco@iaa.es}	
\author[sname='Carpenter']{Michael~Carpenter} \affiliation{Johns Hopkins University Applied Physics Laboratory, 11100 Johns Hopkins Road, Laurel, Maryland, USA}\email{michael.carpenter@jhuapl.edu}	
\author[sname='Bell']{Alexander~Bell} \affiliation{Institut für Sonnenphysik (KIS), Georges-Köhler-Allee 401a, 79110 Freiburg, Germany}\email{albe@leibniz-kis.de}	
\author[orcid=0000-0001-7764-6895,sname='Martínez~Pillet']{Valentín~Martínez~Pillet} \affiliation{Instituto de Astrofísica de Canarias, Vía Láctea, s/n, E-38205 La Laguna, Spain}\affiliation{Spanish Space Solar Physics Consortium}\email{vmpillet@iac.es}
		
\author[orcid=0000-0002-7318-3536,sname='Bailén']{Francisco~Javier~Bailén} \affiliation{Instituto de Astrofísica de Andalucía, CSIC, Glorieta de la Astronomía s/n, 18008 Granada, Spain}\affiliation{Spanish Space Solar Physics Consortium}\email{fbailen@iaa.es}	
\author[orcid=0000-0002-2055-441X,sname='Blanco~Rodríguez']{Julian~Blanco~Rodríguez} \affiliation{Universitat de Valencia Catedrático José Beltrán 2, E-46980 Paterna-Valencia, Spain}\affiliation{Spanish Space Solar Physics Consortium}\email{julian.blanco@uv.es}	
\author[orcid=0000-0003-4319-2009,sname='Castellanos~Durán']{Juan~Sebastián~Castellanos~Durán} \affiliation{Max Planck Institute for Solar System Research, Justus-von-Liebig-Weg 3, 37077 G\"{o}ttingen, Germany}\email{castellanos@mps.mpg.de}	
\author[orcid=0009-0002-6808-5154,sname='Harnes']{Edvarda~Harnes} \affiliation{Max Planck Institute for Solar System Research, Justus-von-Liebig-Weg 3, 37077 G\"{o}ttingen, Germany}\email{harnes@mps.mpg.de}	
\author[orcid=0000-0002-4669-5376,sname='Ishikawa']{Ryohtaroh~T.~Ishikawa} \affiliation{National Institute for Fusion Science, 322-6 Oroshi-cho, Toki City 509-5292, Japan}\email{ishikawa.ryohtaro@nifs.ac.jp}	
\author[orcid=0000-0001-7452-0656,sname='Kawabata']{Yusuke~Kawabata} \affiliation{National Astronomical Observatory of Japan, 2-21-1 Osawa, Mitaka, Tokyo 181-8588, Japan}\email{kawabata.yusuke@nao.ac.jp}	
\author[orcid=0000-0002-1043-9944,sname='Matsumoto']{Takuma~Matsumoto} \affiliation{Centre for Integrated Data Science, Institute for Space-Earth Environmental Research, Nagoya University, Furocho, Chikusa-ku, Nagoya, Aichi 464-8601, Japan}\email{takuma.matsumoto@gmail.com}	
\author[orcid=0000-0002-7044-6281,sname='Oba']{Takayoshi~Oba} \affiliation{Advanced Research Center for Space Science and Technology, Institute of Science and Engineering, Kanazawa University, Kakuma-machi, Kanazawa, Ishikawa 920-1192, Japan}\affiliation{Max Planck Institute for Solar System Research, Justus-von-Liebig-Weg 3, 37077 G\"{o}ttingen, Germany}\email{oba@mps.mpg.de}	
\author[orcid=0000-0003-0175-6232,sname='Siu-Tapia']{Azaymi~L.~Siu-Tapia} \affiliation{Instituto de Astrofísica de Andalucía, CSIC, Glorieta de la Astronomía s/n, 18008 Granada, Spain}\affiliation{Spanish Space Solar Physics Consortium}\email{siu@iaa.es}	
\author[orcid=0000-0003-1483-4535,sname='Strecker']{Hanna~Strecker} \affiliation{Instituto de Astrofísica de Andalucía, CSIC, Glorieta de la Astronomía s/n, 18008 Granada, Spain}\affiliation{Spanish Space Solar Physics Consortium}\email{streckerh@iaa.es}	
\author[orcid=0000-0003-1971-5551,sname='Vukadinović']{Dušan~Vukadinović} \affiliation{Institut für Physik, Universität Graz, Universitätsplatz 5, 8010 Graz, Austria}\affiliation{Max Planck Institute for Solar System Research, Justus-von-Liebig-Weg 3, 37077 G\"{o}ttingen, Germany}\email{vukadinovic@mps.mpg.de}	

\newcommand{\Ipre}{I_{\mathrm{pre}}}
\newcommand{\Ipeak}{I_{\mathrm{peak}}}
\newcommand{\Ppre}{P_{\mathrm{pre}}}
\newcommand{\Ppost}{P_{\mathrm{post}}}
\newcommand{\kms}{\mathrm{km\,s^{-1}}}

\begin{abstract}
Magnetic oscillations permeating the solar atmosphere can be altered by flares. This could be due to changes in the atmospheric magnetic waveguide, the thermodynamic stratification, or both. To investigate this, we examine oscillations in a magnetic pore crossed in part by the ribbon of a C8.2 flare, using observations from the {\sc Sunrise} Ultraviolet Spectropolarimeter and Imager (SUSI) aboard {\sc Sunrise iii}. We track the core intensities of eight spectral lines between 327--329\,nm and show that oscillatory power in the 4.2--7.0\,mHz range, measured around 30 minutes after the flare peak, exceeds the pre-event value in all eight lines by factors of 1.60--8.71. Conversely, in the 7.0--11.5\,mHz band, the post-event intensity power is lower -- approximately 0.13--0.41 of the original power in seven of the observed lines. Line-center velocity measurements support the lower-frequency enhancement. This delayed, line-dependent shift toward the lower-frequency band is consistent with flare-induced changes in wave transmission, reflection, or resonant response, although thermodynamic and magnetic effects cannot yet be separated.
\end{abstract}

\section{Introduction}
\label{sec:introduction}

Magnetohydrodynamic (MHD) waves are an important part of the dynamics and energy balance of the solar atmosphere. They can transport energy from the lower atmosphere and, if sufficiently dissipated, contribute to chromospheric and coronal heating \citep[e.g.,][]{2005LRSP....2....3N, 2015SSRv..190..103J}. Measurements of their amplitudes, frequencies, and phase relations therefore constrain both their contribution to the atmospheric energy budget and the structure and physical conditions of the plasma \citep{2013SSRv..175....1M, 2019mwsa.book.....R}. Sunspots and pores are well suited to such studies because their magnetic fields guide waves from the photosphere toward the upper atmosphere and support oscillations over a broad range of frequencies \citep{2015LRSP...12....6K,2023LRSP...20....1J}.

The oscillatory spectrum measured in a sunspot or pore reflects both the wave power supplied at lower atmospheric levels and the filtering and propagation imposed by the stratified magnetic atmosphere \citep{2006ApJ...640.1153C}. Temperature, density, radiative losses, and magnetic-field inclination influence the effective acoustic cutoff frequency, below which waves cannot normally propagate upward \citep{2004Natur.430..536D, 2006ApJ...648L.151J, 2006ApJ...647L..77M, 2013ApJ...779..168J}. Mode conversion, shock formation, and partial reflection from steep atmospheric gradients can further modify the measured spectrum or produce resonant behavior \citep{2015SSRv..190..103J, 2020NatAs...4..220J, 2021NatAs...5....5J, 2021NatAs...5....2F, 2021A&A...649A.169S, 2022NatCo..13..479S}. Direct observations show that the oscillatory spectrum inside a pore can differ markedly from the surrounding $p$-mode spectrum, with several discrete components and particularly strong power around 4--6\,mHz. Such behavior has been associated with coherent modes and resonant properties of the magnetic waveguide
\citep{2021A&A...649A.169S, 2022ApJ...938..143G, 2024A&A...688A...2J}. 

The measured signal also depends on the spectral line \citep{2008A&A...480..515C, 2012ApJ...749..136L}. Each spectral line samples a finite and generally extended range of atmospheric layers rather than a single geometrical height. Its contribution function, which describes where in the atmosphere the measured radiation originates, and its response functions, which describe how the radiation changes in response to perturbations in atmospheric quantities, depend on the local thermodynamic and velocity structure and can therefore shift or change shape when heating or flows modify the atmosphere \citep{2017SSRv..210..109D, 2014ApJ...795....9F, 2026MNRAS.549g1001C}. Moreover, the visibility of an oscillation in a given line depends on how perturbations in temperature, density, velocity, opacity, and source function are mapped into changes in its line-core intensity or Doppler shift. Consequently, spectral lines with similar or overlapping estimated formation ranges can emphasize different components of the same underlying wave field and may show different dominant frequencies \citep{2026ApJ..1005L..69J}. Thus, different lines are not expected to respond in the same way to a rapid atmospheric change.

This dependence of both wave propagation and spectral-line formation on the atmospheric state makes flares a natural setting in which to search for rapid changes in the measured oscillatory spectrum \citep{2024ARA&A..62..437F}. Solar flares rapidly convert stored magnetic energy into plasma heating, particle acceleration, mass motions, radiation, and waves \citep{2008ApJ...675.1645F, 2011SSRv..159...19F, 2017LRSP...14....2B}. They may affect pre-existing oscillations by changing the temperature and density structure, producing flows \citep{2025A&A...699A.121K}, altering the magnetic geometry, or changing the effective reflecting layers of the atmosphere. In particular, changes in temperature modify the sound speed, while changes in magnetic-field inclination alter the effective cutoff for field-guided magnetoacoustic waves. Flare-driven heating or magnetic reconfiguration can therefore shift the range of frequencies able to propagate through the atmosphere, producing frequency-dependent enhancement or suppression without necessarily generating a new wave mode \citep{2021MNRAS.503.2444M,2024ApJ...961..231W}. 

Oscillations can also be modified by processes other than flares. For example, an impulsive downflow above a sunspot was found to excite a chromospheric oscillation whose period increased from approximately 2.7 to 3.3 minutes, interpreted as the response of the gravitationally stratified atmosphere to the impulsive disturbance \citep{2016ApJ...821L..30K}. The causal relation may also operate in the opposite direction: \citet{2009A&A...505..791S} reported an increase in three-minute sunspot oscillations before several flares and proposed that waves traveling from the sunspot could modulate or help initiate the energy release. At coronal heights, flare-related disturbances have been associated with large-scale Doppler and transverse oscillations \citep{2011A&A...528L..12T}, propagating fast wave trains \citep{2013A&A...554A.144Y}, and long-period oscillations in flaring loops \citep{2016SoPh..291.3289C}. Related flare--oscillation coupling has also been examined on larger spatial scales. \citet{2015MNRAS.450..956B} searched for global stellar oscillations that might be excited by flares, while \citet{2026ApJ..1004..212L} studied how solar flares affect decayless kink oscillations in nearby coronal loops.

Observations of the lower solar atmosphere have revealed several forms of flare-related oscillatory response. \citet{2007ApJ...670L.147K} detected high-frequency chromospheric oscillations following a large flare and interpreted them as a possible flare-excited response. Enhanced three-minute signals have also been found in flare-related Ly$\alpha$ emission \citep{2017ApJ...848L...8M} and in spatially resolved ultraviolet observations near hard X-ray sources \citep{2020ApJ...903...19F}. Using H$\alpha$ and Ca~{\sc ii}~854.2\,nm observations, \citet{2021MNRAS.503.2444M} found that both the locations and characteristic periods of chromospheric oscillations changed after a flare. They interpreted the generally longer periods as evidence of changes in the magnetic configuration, supported by changes in the connectivity of coronal loops. In their interpretation, the change in field inclination modified the effective acoustic cutoff and therefore the periods able to propagate. \citet{2024MNRAS.527.5916M} later showed that the intensity and line-of-sight velocity responses do not necessarily occur together and reported a velocity signature well after the impulsive phase. From ultraviolet imaging, \citet{2024ApJ...961..231W} identified an enhanced three-minute oscillation above a sunspot during a flare, together with a modest change in the inferred magnetic-field inclination. These studies show that the response can include excitation, suppression, frequency changes, changes in the spatial distribution of oscillatory power, and signatures that appear only after the impulsive flare phase. 

Most previous measurements relied on imaging data or on a small number of spectral lines. It is therefore difficult to determine whether a measured power change represents a broad modification of the lower atmosphere or the particular response of one line. It also remains unclear whether a frequency-dependent change can emerge only after the impulsive flare phase and be detected simultaneously in several lower-atmospheric spectral lines. This distinction becomes especially important during a flare, when line profiles can brighten, become centrally reversed, or change from absorption to emission. A measurement at a fixed wavelength can then combine true intensity changes with Doppler shifts and changes in the line shape. Following the fitted central wavelengths of several lines simultaneously provides a more reliable comparison. It also avoids assigning each line to a single, sharply defined formation height; instead, the different and partly overlapping atmospheric responses and perturbation sensitivities of the lines are treated as part of the measurement.

The \textsc{Sunrise} Ultraviolet Spectropolarimeter and Imager (SUSI; \citealt{2025SoPh..300...65F,2025SoPh..300...58I}) aboard the third flight of the \textsc{Sunrise} balloon-borne solar observatory \citep{2025SoPh..300...75K,2026ApJ..1005L..64S} provides such measurements in the crowded near-ultraviolet spectrum. Its seeing-free observations, following the successful first two \textsc{Sunrise} flights \citep{2010ApJ...723L.127S,2011SoPh..268....1B,2017ApJS..229....2S}, combine high spatial resolution, repeated rapid raster scans, and many spectral lines recorded at the same locations and times. The present work uses a particular repeated-raster time series obtained by SUSI while continuously sampling the same active-region field of view. Different parts and aspects of this same observing sequence have been used to characterize line-dependent oscillations in the sunspot umbra, map oscillatory power across an active-region transect, and identify MHD wave modes associated with the boundary of the pore during the flare-free part of the sequence \citep{2026ApJ..1005L..69J,2026ApJ..1005L..65J,2026ApJ..1005L..81J}.

Here, we investigate how the pore responded to a compact C8.2 flare whose ribbon crossed part of the pore during the first part of this SUSI time series. We follow the fitted line-core intensities of eight near-ultraviolet lines and compare three intervals of similar duration: a pre-event reference, an early post-event interval, and a late post-event interval. Refined global wavelet spectra \citep{2025NRvMP...5...21J} are used to identify statistically significant oscillatory power, while the fitted line-of-sight velocities provide supporting information from a different observable. We examine how the significant frequency spectrum changed after the flare, whether the response was present in the early post-event interval or developed later, and how it differed among spectral lines with different atmospheric sensitivities.

\section{Observations}
\label{sec:observations}

We analyze a subset of the {\sc Sunrise iii} observations of NOAA active region 13744 obtained on 2024 July 14, when the target was close to disk center ($\mu\approx0.96$). SUSI sampled the 327--329\,nm wavelength range with a spectral sampling of approximately 0.83\,pm pixel$^{-1}$ and a spatial sampling of $\approx0\farcs03$ pixel$^{-1}$ along the slit \citep{2025SoPh..300...65F}. A narrow raster covered the pore and its immediate surroundings with a cadence of about 39\,s. The full observing sequence contains 180 rasters acquired between 21:40 and 23:39\,UTC. We use the first 81 rasters, from $t=0$ to 3144.8\,s, which include the pre-event atmosphere, the complete flare-related brightening, and the subsequent post-event evolution. The flare ribbon crossed the selected pore, including part of the region in which the oscillations were analyzed.

The pointing system, consisting of the gondola \citep{2025SoPh..300..112B}  and the Correlating Wavefront Sensor \citep{2026SoPh..301...57B}, ensured the high stability of the observations, of particular importance for the study of timeseries.

The event coincides with a C8.2 flare assigned by the SWPC/HEK catalog to NOAA active region 13744. The cataloged event started at 21:54\,UTC, peaked at 21:58\,UTC, and ended at 22:02\,UTC. A small contemporaneous increase is present in the full-Sun GOES/XRS irradiances, while the spatially coincident response in the SDO/AIA images \citep{2012SoPh..275....3P,2012SoPh..275...17L} supports the catalog association with the SUSI target (Appendix~\ref{app:aia-goes}; Figure~\ref{fig:aia-goes-context}). The event is visible in all six AIA passbands examined here, with spatially extended and locally saturated emission in the EUV images. During a flare, emission recorded by these broad, multithermal passbands can originate from plasma at different temperatures and atmospheric locations, including flare-ribbon plasma in the lower atmosphere. We therefore use the AIA images to establish the spatial and temporal context of the event, but do not infer a plasma temperature from their intensities alone.

\begin{figure*}[t]
\centering
\includegraphics[width=\textwidth]{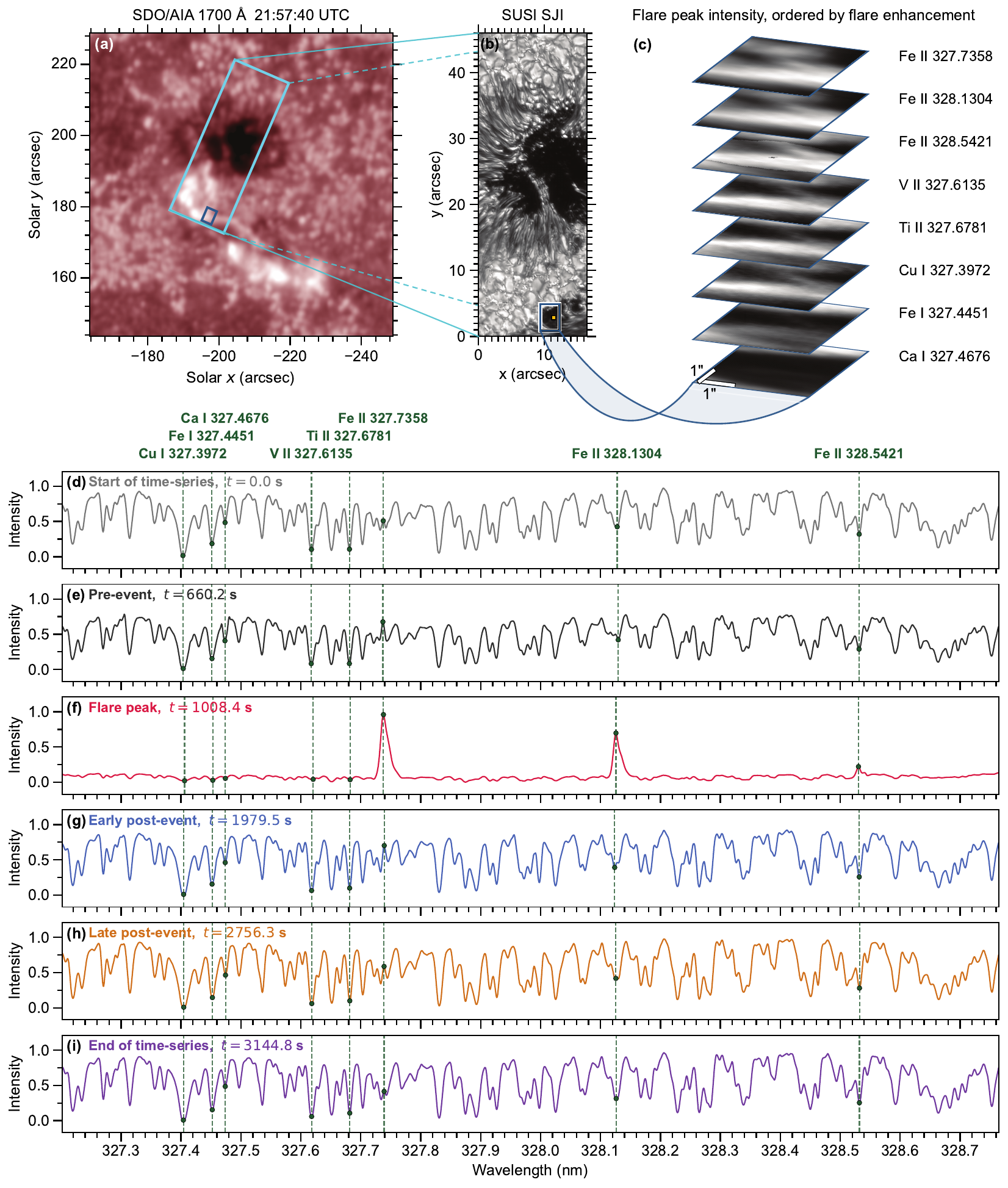}
\caption{Observational context and spectral evolution. (a) SDO/AIA 1700\,\AA\ image near the flare peak, with the SUSI and pore field of view marked. (b) SUSI 325\,nm slit-jaw image at 21:57:39.9\,UTC; the box encloses the pore region analyzed here and the yellow symbol identifies a spectral-sampling pixel within the strong-response region. (c) Tracked line-core intensity maps at 21:57:39.9\,UTC, stacked from the weakest to the strongest median flare enhancement. (d)--(i) Normalized spectra at the marked pixel in panel (b) at the sequence start, pre-event interval, flare peak, early and late post-event intervals, and sequence end. Vertical markers identify the eight analyzed lines. Line centers are determined from these processed spectra, while intensities are sampled on the original scale. The \href{https://WaLSA.team/sunrise/figure1_pore_flare_animation}{online animation} follows the spectrum throughout the 52-minute sequence.}
\label{fig:context}
\end{figure*}

We use only the calibrated Stokes $I$ spectra because the corresponding full-Stokes profiles were not yet available at this stage of the {\sc Sunrise iii}/SUSI program. The standard reduction included dark- and flat-field corrections, wavelength calibration, and instrumental corrections. The time sequence was subsequently corrected for solar rotation, field rotation, and residual image motion, following the processing described in earlier studies of these observations \citep{2026ApJ..1005L..69J, 2026ApJ..1005L..65J, 2026ApJ..1005L..81J}. 
Figure~\ref{fig:context} shows the observing geometry, the multiline flare response, and representative spectra throughout the sequence. The SUSI slit-jaw image at 325\,nm provides the local context, while an SDO/AIA 1700\,\AA\ image places the small SUSI field of view within the active region. 

We retain eight lines that are sufficiently isolated for stable tracking and display a range of flare responses: Cu~{\sc i}~327.3972\,nm, Fe~{\sc i}~327.4451\,nm, Ca~{\sc i}~327.4676\,nm, V~{\sc ii}~327.6135\,nm, Ti~{\sc ii}~327.6781\,nm, Fe~{\sc ii}~327.7358\,nm, Fe~{\sc ii}~328.1304\,nm, and Fe~{\sc ii}~328.5421\,nm. This is deliberately a limited selection for an early-science Letter. The near-ultraviolet spectrum contains many additional usable transitions \citep[see Figure~1 of][]{2026ApJ..1005L..69J}, but their complete interpretation requires line-by-line response functions and non-LTE modeling and is left to follow-up work.

\section{Analysis and Results}
\label{sec:analysis-results}

Our analysis has two connected parts. We first track the eight selected line cores throughout the pore and use their intensity changes to identify the part of the pore with the strongest multiline brightening during the flare. The AIA images show that the flare ribbon crossed the selected pore region. The strong-response subset used for the oscillation analysis is not defined from the AIA ribbon morphology; instead, its pixels are selected exclusively from the SUSI multiline line-core response (Section~\ref{subsec:intervals}). Thus, this independent selection refers to the method used to define the mask, rather than to the spatial separation from the flare ribbon. We note that the strong-response mask overlaps with the part of the pore crossed by the AIA flare ribbon. We then use the tracked line-core intensity time series as the primary observable to compare the oscillatory state of this region during equal-duration pre-event, early post-event, and late post-event intervals, examining both the frequency-resolved fraction of pixels with significant oscillatory power and the corresponding band-integrated power. Line-center velocities provide a secondary check, using a different observable, of the main post-event frequency band. 

\subsection{Fitting spectral lines}
\label{subsec:fitting}

The spectrum is crowded, and a common, uncontaminated nearby continuum reference cannot be defined across all eight lines. Fixed-wavelength measurements are also sensitive to Doppler shifts and changes in line-profile shape. We therefore determine the line center separately for every spatial pixel and raster. To suppress pixel-scale noise before locating the extrema, we apply mild $3\times3$ spatial boxcar averaging and local three-pixel spectral smoothing, following \citet{2026ApJ..1005L..65J}. These operations introduce spatial smoothing comparable to the diffraction limit and spectral broadening below the instrumental resolution. The smoothed spectrum is then normalized to its own intensity range only for identifying the line centers.

We first used the \texttt{LineFit} routine \citep{2026FrASS..1343842J, walsalib..2026...23043079} from the WaLSAlib repository\footnote{\url{https://github.com/WaLSAteam/WaLSAlib}} on representative spectra from different spatial positions and observing times. These included the pre-event atmosphere, the flare peak, and the transition back to absorption. \texttt{LineFit} does not require symmetric line profiles: it uses bounded Voigt-family fits, including an asymmetric-Voigt option, and incorporates morphology-aware treatment of asymmetric, blended, and split-core profiles. We used the asymmetric-Voigt option in these tests to establish suitable search windows and profile treatments for each transition and to verify that the intended spectral feature was identified. 

For efficient application to the complete spatial and temporal cube, the validated search windows were subsequently used by a five-point quadratic core tracker anchored to the calibrated reference wavelength of each transition. The three Fe~{\sc ii} lines were allowed to switch locally among absorption, central-reversal, and full-emission treatments according to their profile shape. We inspected spectra from many spatial positions and times, including cases near transitions among the adopted profile treatments, to verify that the tracker continued to follow the intended central feature. Finally, the line-core intensity was sampled from the unnormalized, original-scale spectral cube at the tracked wavelength. Thus, neither the smoothing nor the normalization used to locate the line center enters the measured intensity.

The tracked centers also provide line-of-sight velocity estimates. These measurements are less secure during strong emission or rapid changes in profile shape, and intensity is therefore the main observable in this Letter. We use velocity only as a secondary check of the principal post-event frequency response.

Near the flare peak, the Fe~{\sc ii} profiles also show an asymmetric red-wing component superposed on the brighter main profile. Similar components in IRIS Fe~{\sc ii} and other singly ionized lines have been interpreted as emission from rapidly downflowing chromospheric-condensation plasma \citep[e.g.,][]{2020ApJ...895....6G}. In the SUSI profiles inspected here, the distinct red-wing component is confined mainly to the flare interval excluded from the oscillation analysis; no comparable separate component is evident during the early or late post-event intervals. Near the flare peak, however, this component can influence both the tracked line center and the intensity sampled at that center. The reported peak-intensity ratios should therefore be understood as enhancements of the selected tracked core feature, rather than measurements of the wavelength-integrated line emission. The flare frames are excluded from the oscillation analysis, limiting their influence on the pre-event and post-event intensity comparison. The effect is particularly important for line-center velocities because any displacement of the selected component is interpreted directly as a Doppler shift. We therefore do not interpret flare-time velocities and use the post-event velocity analysis only as a secondary check.

\subsection{Event intervals and strong-response mask}
\label{subsec:intervals}

The pre-event reference is the median of frames 0--17, where each frame represents one completed raster. The recorded time of each raster corresponds to the beginning of its left-to-right scan. For the pixel-by-pixel time-series analysis, we add the position-dependent scan offset derived from the raster cadence and the horizontal scan coordinate.

The pre-event interval comprises frames 0--17 and ends before the onset of the sustained flare-related multiline brightening in the SUSI line-core intensities. The flare-related interval covers frames 18--40 and includes the main brightening and its subsequent decay; the spatially averaged intensity reaches its maximum at frame 26 ($t=1008.4$\,s from the beginning of the sequence; 21:57:39.9\,UTC). These flare-related frames are excluded from the oscillation analysis because the large, nonperiodic intensity excursion can introduce broad wavelet power even in the absence of a periodic signal.

For the oscillation comparison, we use three nonoverlapping intervals of identical duration: pre-event (frames 0--17; $t=0$--$661$\,s), early post-event (frames 41--58; $t=1631$--$2291$\,s), and late post-event (frames 63--80; $t=2485$--$3145$\,s). Each interval contains 18 rasters, corresponding to approximately 11 minutes, thereby providing the same frequency sampling and wavelet's cone-of-influence limitations for all three comparisons. The early post-event interval begins immediately after the flare-related interval. Frames 59--62 are retained as a short separation, after which the late post-event interval uses the final 18 rasters of the analyzed sequence.

\begin{figure}[t]
\centering
\includegraphics[width=0.55\columnwidth]{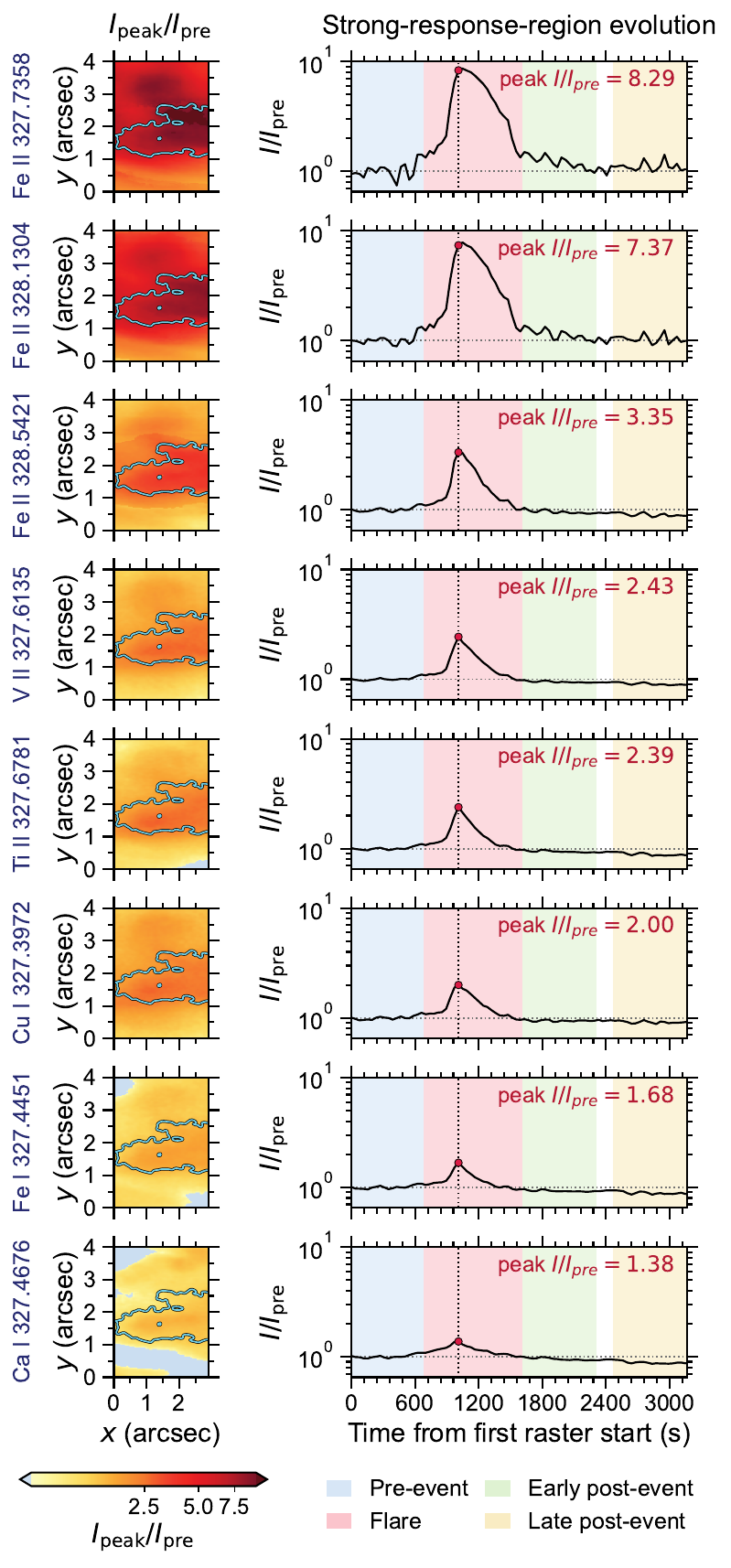}
\caption{Direct multiline response to the flare, with the spectral lines ordered by their median peak enhancement within the strong-response region. For each line, the left-hand panel shows the pixel-by-pixel ratio $\Ipeak/\Ipre$. All ratio maps use the same color scale. The cyan contour marks the boundary of the same composite strong-response region in every map. The right-hand panel in each row shows the temporal evolution of the median $I/\Ipre$ within this region. The background colors identify the pre-event, flare, early post-event, and late post-event intervals, while the vertical dotted line marks the intensity peak at $t=1008.4$\,s. The first raster began at 21:40:51.5\,UTC.}
\label{fig:flare}
\end{figure}

We next define a subset of pixels with a strong multiline brightening during the flare, hereafter the strong-response mask. For every spectral line and spatial pixel, we calculate the local pre-event median, $\Ipre$, and normalize the intensity as $I/\Ipre$. This normalization allows the relative brightening to be compared among lines with different absolute intensities. 
For the detection threshold, we quantify the temporal pre-event variability separately for every spectral line and spatial pixel using the scaled median absolute deviation, $\sigma_{\rm pre}=1.4826\,{\rm median}\left|I(t)-\Ipre\right|$, where both $\Ipre$ and the median absolute deviation are calculated from the same 18 pre-event rasters. The factor 1.4826 converts the median absolute deviation to the corresponding standard deviation for normally distributed fluctuations \citep{1993JASA...88.1273R}. Across the selected pore region, the line-by-line median values of $\sigma_{\rm pre}/\Ipre$ range from 3.3\% to 12.9\%, with a median of 5.4\% over all retained line--pixel combinations. A pixel is classified as strongly brightened in a given line when $(I_{\rm peak}-\Ipre)/\sigma_{\rm pre}\geq5$. The strong-response mask comprises pixels that satisfy this criterion in all eight lines and in which all three Fe~{\sc ii} lines are classified as being in full emission at the flare peak. It contains 2081 pixels, corresponding to 26.5\% of the selected pore region. The cyan contour in the left column of Figure~\ref{fig:flare} marks the common strong-response region. We note that it is not expected to follow the brightest structure in every spectral line. Small differences can arise from the opacity and source-function response of each line, Doppler shifts, broad and overlapping formation ranges, and possible horizontal displacements among structures sampled at different atmospheric levels. The event is therefore clearly detected beyond the strongly emitting Fe~{\sc ii} lines, but its measured strength and morphology depend substantially on the selected transition.

\subsection{Oscillation analysis}
\label{subsec:wavelets}

Each intensity time series is linearly detrended and apodized with a tapered-cosine (Tukey) window (with a tapered fraction $\alpha=0.1$) before the wavelet calculation with WaLSAtools\footnote{\url{https://github.com/WaLSAteam/WaLSAtools}} \citep{2025NRvMP...5...21J,walsatools..2025...17569951}. We calculate the wavelet transform using a Morlet mother wavelet \citep{1998BAMS...79...61T, 2016ApJ...825..110A}. Statistical significance is assessed separately for every line, pixel, and interval at the 95\% level using 1000 randomized surrogate time series generated with the same WaLSAtools procedure; no parametric noise model is assumed. The refined global wavelet spectrum (RGWS) is constructed by integrating over time only those wavelet-power measurements that pass the 95\% statistical significance criterion and lie outside the cone-of-influence region affected by edge effects. Thus, RGWS power throughout this paper already refers to statistically selected wavelet power. Because each event interval contains only 18 rasters, the longest period for which any samples remain outside the cone of influence is approximately 240\,s. On the wavelet-frequency grid, the first usable frequency is 4.213\,mHz. We therefore set 4.2\,mHz as the lower boundary of the interval-based analysis. The absence of RGWS values below 4.2\,mHz is imposed by the finite interval length and the cone-of-influence restriction.

We describe the oscillatory response using two complementary quantities. First, at each frequency we define the frequency-resolved RGWS detection fraction as the percentage of pixels in the strong-response mask whose RGWS contains power passing the 95\% significance threshold at that frequency. This quantity measures how commonly statistically selected power is detected within the region at an individual frequency; it does not measure the magnitude of the RGWS power. The same spatial mask, significance criterion, frequency grid, and analysis settings are used for every spectral line and event interval. No normalization to the maximum of an individual line or interval is applied.

Second, we integrate the RGWS power over the 4.2--7.0\,mHz (periods of $\approx$2.4--4.0\,min encompassing the classical three-minute chromospheric oscillation) and 7.0--11.5\,mHz ($\approx$1.5--2.4\,min) bands and calculate separate early- and late-post-event power ratios, $P_{\rm early}/P_{\rm pre}$ and $P_{\rm late}/P_{\rm pre}$. The lower frequency band contains the main pore oscillations measured here, whereas the upper band tests the higher-frequency response. 

Using these same frequency bands, we also calculate a band-coverage fraction: the percentage of pixels with at least one RGWS detection anywhere within a given band, with each pixel counted only once per band. This is a band-level summary of the detections and is distinct from the frequency-resolved detection fraction, which is evaluated separately at every frequency and displayed in Figure~\ref{fig:rgws-occurrence}. We use the band-coverage fraction for the spatial comparison between the strong-response mask and the remainder of the selected pore region.

To account for the spatial correlation between neighboring pixels, we divide the region into nonoverlapping $5\times5$-pixel blocks. For each bootstrap realization, these spatial blocks are resampled with replacement, and the post-event to pre-event ratio of the region-mean power is recalculated. The 2.5th and 97.5th percentiles of 5000 paired realizations define the 95\% bootstrap confidence interval. This interval therefore estimates the uncertainty while preserving the local spatial correlations within each block.

The selected spectral lines are expected to have broad and overlapping response functions, particularly in the dynamic flare atmosphere. We therefore compare them as distinct but overlapping spectral responses and do not impose a strict ordering by formation height.

As a threshold-independent robustness test that also avoids line-center tracking and Fe~{\sc ii} profile-state classification, we calculate unthresholded Fourier band powers from original-scale intensities averaged over fixed, symmetric windows spanning 17 wavelength samples ($\pm6.7$\,pm) windows centered on the calibrated reference wavelength of each line. The short 18-raster intervals contain too few cycles for a systematic eight-line phase-delay analysis, which is left to longer-sequence work.

\subsection{A multiline flare response}
\label{subsec:flare-response}

All eight line cores brighten during the event, although the amplitudes and spatial patterns differ strongly (Figure~\ref{fig:flare}). The most pronounced changes occur in the Fe~{\sc ii} lines, which develop full-emission profiles over much of the strong-response region. Their median peak ratios are $\Ipeak/\Ipre=8.29$ for Fe~{\sc ii}~327.7358\,nm, 7.37 for Fe~{\sc ii}~328.1304\,nm, and 3.35 for Fe~{\sc ii}~328.5421\,nm.
The other five lines remain mainly in absorption, but their cores fill in substantially. Their corresponding peak ratios are 2.43 for V~{\sc ii}, 2.39 for Ti~{\sc ii}, 2.00 for Cu~{\sc i}, 1.68 for Fe~{\sc i}, and 1.38 for Ca~{\sc i}. These ratios measure the enhancement of the tracked line-core intensity. In the Fe~{\sc ii} lines, this includes the development of central reversals and, over part of the strong-response mask, full emission. The remaining lines mainly show core filling and brightening rather than the same full-emission behavior. The tracked Cu~{\sc i} response is particularly clear compared with a fixed-wavelength image, in which Doppler motion can hide much of the brightening.

\subsection{Frequency-resolved changes in the RGWS detection fraction}
\label{subsec:occurrence}

\begin{figure*}[t]
\centering
\includegraphics[width=\textwidth]{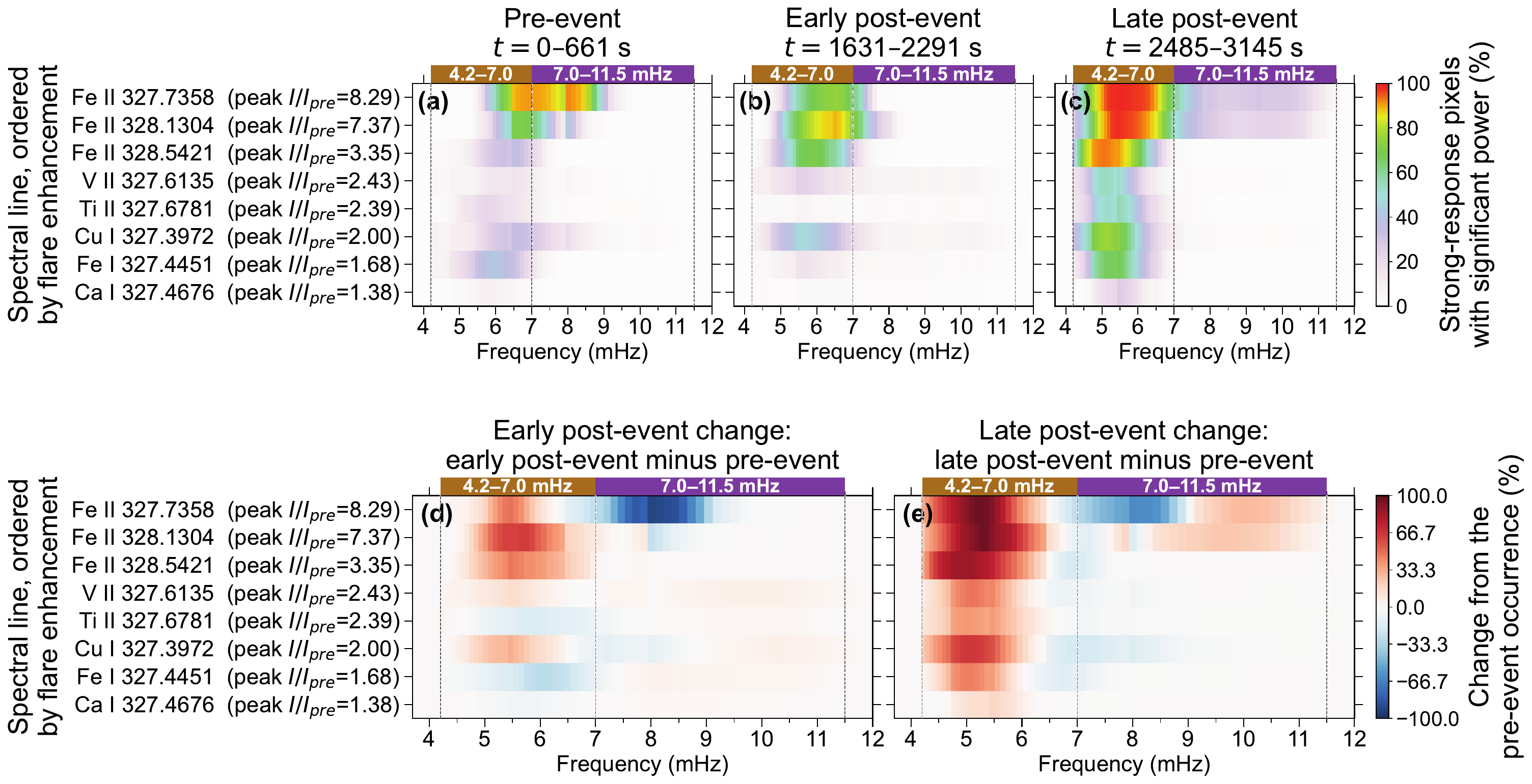}
\caption{Frequency-resolved RGWS detection fraction within the strong-response mask. The spectral lines are ordered by their line-core enhancement at the flare peak. Panels (a)--(c) show the absolute percentage of pixels within the same strong-response mask that have an RGWS detection at each frequency during the pre-event, early post-event, and late post-event intervals. At each frequency, a detection requires the wavelet power to pass the 95\% significance test and the cone-of-influence restriction. No separate normalization is applied to any line or interval, and all three panels use the same 0--100\% scale. Panels (d) and (e) show the early-minus-pre and late-minus-pre differences in percentage points. The 4.2--7.0 and 7.0--11.5\,mHz bands are marked in all panels. The displayed quantity is the frequency-resolved RGWS detection fraction, not the magnitude of the RGWS power or the band-coverage fraction used for the spatial comparison. The sharp lower boundary at 4.2\,mHz is set by the duration of the 18-raster intervals and the cone-of-influence restriction, rather than by an observed absence of lower-frequency oscillations.}
\label{fig:rgws-occurrence}
\end{figure*}

The pore already contains oscillatory power before the flare. The post-event behavior therefore represents a change to the existing oscillatory state rather than the first appearance of oscillations. Figure~\ref{fig:rgws-occurrence} shows, for every spectral line and frequency, the percentage of pixels within the fixed strong-response mask that have an RGWS detection. We refer to this quantity as the frequency-resolved RGWS detection fraction. These are absolute percentages on a common 0--100\% scale and are not normalized separately for each line or event interval, so the pre-event, early post-event, and late post-event measurements can be compared directly. The pre-event frequency distributions differ among the lines, consistent with their different and overlapping atmospheric responses.

The early post-event response is strongly line dependent. In the 4.2--7.0\,mHz band, the band-averaged detection fraction increases for Fe~{\sc ii}~327.7358\,nm, Fe~{\sc ii}~328.1304\,nm, Fe~{\sc ii}~328.5421\,nm, V~{\sc ii}, and Cu~{\sc i}, but decreases for Ti~{\sc ii}, Fe~{\sc i}, and Ca~{\sc i}. In the 7.0--11.5\,mHz band, Fe~{\sc ii}~327.7358\,nm shows the clearest early reduction; Fe~{\sc ii}~328.1304\,nm and Cu~{\sc i} show only small net decreases, whereas the other five lines show smaller positive changes. A measurement from any single line would therefore provide an incomplete account of the early post-event oscillatory response.

A clearer common pattern appears later. By the late post-event interval, the band-averaged 4.2--7.0\,mHz detection fraction is greater than its pre-event value in all eight lines. In the 7.0--11.5\,mHz band, it is lower in six lines and effectively unchanged for Ca~{\sc i}; Fe~{\sc ii}~328.1304\,nm is the only line showing a clear overall increase. The frequency-resolved measurements nevertheless reveal structure within this broad band. Fe~{\sc ii}~327.7358\,nm shows a negative late-minus-pre difference mainly between 7 and 9\,mHz, but a positive difference at approximately 9--11.5\,mHz. Conversely, Fe~{\sc ii}~328.1304\,nm shows a modest decrease toward the lower end of the band before its higher-frequency increase dominates the band average. These narrower features are combined in the band-averaged values reported in Table~\ref{tab:detection-change}. We refer to this mixed early response, followed by a common lower-frequency increase and a predominantly reduced higher-frequency response, as a delayed reorganization of the detected oscillatory power. For the numerical summaries in Table~\ref{tab:detection-change}, we average the frequency-resolved detection fraction over each band by interpolating to the exact band boundaries, applying trapezoidal integration, and dividing by the band width.

\setlength{\tabcolsep}{7.5pt}
\begin{deluxetable}{lcccc}[b]
\tabletypesize{\footnotesize}
\tablewidth{0pt}
\tablecaption{Post-event minus pre-event change in the band-averaged percentage of strong-response pixels with significant RGWS detections
\label{tab:detection-change}}
\tablehead{
\colhead{Spectral line} &
\multicolumn{2}{c}{Early $-$ pre} &
\multicolumn{2}{c}{Late $-$ pre} \\ 
\cmidrule(lr){2-3}\cmidrule(lr){4-5}
\noalign{\vskip-1.6ex}
\colhead{} &
\colhead{4.2--7.0} &
\colhead{7.0--11.5} &
\colhead{4.2--7.0} &
\colhead{7.0--11.5}
}
\startdata
Fe~{\sc ii}~327.7358 & $+11.3$ & $-32.8$ & $+48.9$ & $-10.2$ \\
Fe~{\sc ii}~328.1304 & $+33.5$ & $-1.9$  & $+54.7$ & $+11.2$ \\
Fe~{\sc ii}~328.5421 & $+26.0$ & $+0.4$  & $+46.3$ & $-1.5$  \\
V~{\sc ii}~327.6135  & $+6.2$  & $+3.8$  & $+20.0$ & $-1.3$  \\
Ti~{\sc ii}~327.6781 & $-8.8$  & $+0.6$  & $+20.0$ & $-1.1$  \\
Cu~{\sc i}~327.3972  & $+12.5$ & $-0.8$  & $+29.2$ & $-6.2$  \\
Fe~{\sc i}~327.4451  & $-15.6$ & $+2.1$  & $+14.5$ & $-1.0$  \\
Ca~{\sc i}~327.4676  & $-2.5$  & $+1.3$  & $+6.6$  & $0.0$
\enddata
\tablecomments{Wavelengths are in nm, and frequency bands are in mHz. At each frequency, the RGWS detection percentage is the percentage of pixels in the strong-response mask whose wavelet power passes the 95\% significance criterion and lies outside the cone of influence. For each frequency band, this percentage is averaged over frequency as described in Section~\ref{subsec:occurrence}. Each table entry is the post-event value minus the corresponding pre-event value, expressed in percentage points; for example, $+11.3$ denotes an increase of 11.3 percentage points, not a relative increase of 11.3\%. These values are not band-coverage fractions.}
\end{deluxetable}

As a spatial comparison, we repeated the band-coverage measurement defined in
Section~\ref{subsec:wavelets} over the remaining 5779 pixels of the selected pore region outside the strong-response mask. In the late interval, the 4.2--7.0\,mHz band-coverage fraction increases in six of the eight lines outside the mask, compared with all eight inside it. The two exceptions outside the mask are Fe~{\sc i} and Ca~{\sc i}; Ti~{\sc ii} shows only a small increase. The corresponding late-to-pre-event band-power ratio exceeds unity in six lines outside the mask, with 95\% bootstrap confidence intervals lying entirely above unity for five of them; all eight ratios inside the mask exceed unity. In the 7.0--11.5\,mHz band, the band-coverage fraction decreases in all eight lines in both areas. This oscillatory reorganization therefore extends across the pore, but its lower-frequency component is more consistent among the lines where the flare brightening is strongest. We use the remaining pore as a comparison area, not an unaffected control, because the event may have perturbed the pore more broadly.

\subsection{Changes in RGWS band power}
\label{subsec:band-power}

\begin{figure*}[t]
\centering
\includegraphics[width=\textwidth]{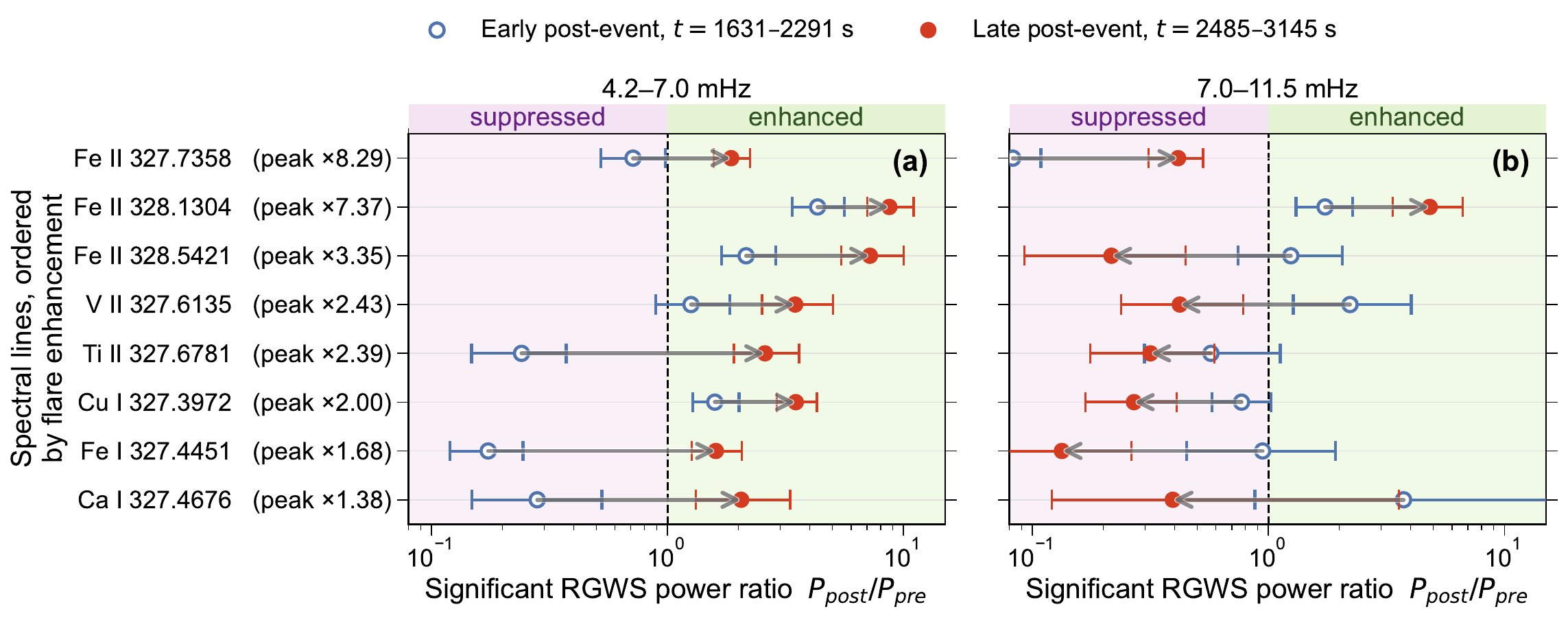}
\caption{RGWS power relative to the pre-event interval. The spectral lines are ordered by their line-core enhancement at the flare peak. Panels (a) and (b) show the 4.2--7.0 and 7.0--11.5\,mHz bands, respectively. Open blue circles denote the early post-event ratio, and filled red circles denote the late post-event ratio; gray connectors show the change between these intervals. Horizontal bars give paired 95\% bootstrap confidence intervals for the ratios of the spatially averaged band powers, obtained by resampling spatial blocks. The vertical line at unity denotes no change from the pre-event state.}
\label{fig:band-power}
\end{figure*}

The preceding section quantified how commonly RGWS oscillations are detected within the strong-response region. We now examine how their band-integrated power changes. The spatially averaged RGWS power depends on both the retained RGWS power and the area over which that power passes the significance test. It is therefore complementary to, but not independent of, the detection fraction. For each post-event interval, Figure~\ref{fig:band-power} compares this quantity with its pre-event value: $\Ppost/\Ppre>1$ denotes enhanced RGWS power, whereas a ratio below unity denotes reduced power.

During the early post-event interval, the 4.2--7.0\,mHz power increases clearly for Fe~{\sc ii}~328.1304\,nm, Fe~{\sc ii}~328.5421\,nm, and Cu~{\sc i}, but decreases for Fe~{\sc ii}~327.7358\,nm, Ti~{\sc ii}, Fe~{\sc i}, and Ca~{\sc i}. The change in V~{\sc ii} is not statistically distinguishable
from no change. In the 7.0--11.5\,mHz band, Fe~{\sc ii}~327.7358\,nm shows a clear reduction, whereas V~{\sc ii} and Fe~{\sc ii}~328.1304\,nm show clear increases. Ratios for Fe~{\sc ii}~328.5421\,nm, Ti~{\sc ii}, Cu~{\sc i}, Fe~{\sc i}, and Ca~{\sc i} are not statistically distinguishable from unity. The corresponding ratios are listed in Table~\ref{tab:band-power-ratios}.

\setlength{\tabcolsep}{8.0pt}
\begin{deluxetable}{lcccc}[b]
\tabletypesize{\footnotesize}
\tablewidth{0pt}
\tablecaption{Post-event/pre-event ratios of spatially averaged RGWS band power
\label{tab:band-power-ratios}}
\tablehead{
\colhead{Spectral line} &
\multicolumn{2}{c}{Early post-event} &
\multicolumn{2}{c}{Late post-event} \\
\cmidrule(lr){2-3}\cmidrule(lr){4-5}
\noalign{\vskip-1.6ex}
\colhead{} &
\colhead{4.2--7.0} &
\colhead{7.0--11.5} &
\colhead{4.2--7.0} &
\colhead{7.0--11.5}
}
\startdata
Fe~{\sc ii}~327.7358 & 0.71 & 0.08 & 1.86 & 0.41 \\
Fe~{\sc ii}~328.1304 & 4.32 & 1.74 & 8.71 & 4.83 \\
Fe~{\sc ii}~328.5421 & 2.15 & 1.24 & 7.20 & 0.22 \\
V~{\sc ii}~327.6135  & 1.26 & 2.22 & 3.47 & 0.42 \\
Ti~{\sc ii}~327.6781 & 0.24 & 0.57 & 2.59 & 0.32 \\
Cu~{\sc i}~327.3972  & 1.59 & 0.77 & 3.50 & 0.27 \\
Fe~{\sc i}~327.4451  & 0.17 & 0.95 & 1.60 & 0.13 \\
Ca~{\sc i}~327.4676  & 0.28 & 3.74 & 2.05 & 0.39
\enddata
\tablecomments{Wavelengths are in nm, and frequency bands are in mHz. Each entry is the ratio of the spatially averaged, band-integrated RGWS power in the stated post-event interval to the corresponding pre-event value. A ratio of 1 denotes no change; values above 1 indicate enhancement and values below 1 indicate reduction. For example, a ratio of 1.86 corresponds to an 86\% increase, whereas a ratio of 0.41 corresponds to a 59\% reduction. The paired 95\% bootstrap confidence intervals are shown in Figure~\ref{fig:band-power}.}
\end{deluxetable}

The late post-event interval displays a much more uniform change in the main pore band. The 4.2--7.0\,mHz power exceeds its pre-event value in all eight lines, with all bootstrap confidence intervals lying above unity. In the 7.0--11.5\,mHz band, the ratio is below unity in seven lines. The reductions are statistically clear for Fe~{\sc ii}~327.7358\,nm, Fe~{\sc ii}~328.5421\,nm, V~{\sc ii}, Ti~{\sc ii}, Cu~{\sc i}, and Fe~{\sc i}. The Ca~{\sc i} confidence interval includes unity, so its individual reduction is not statistically secure. Fe~{\sc ii}~328.1304\,nm is the clear exception, showing enhanced higher-frequency power. The corresponding ratios are given in Table~\ref{tab:band-power-ratios}.

The detection-fraction and band-power measurements therefore describe different aspects of the oscillatory response. Higher-frequency RGWS detections become less widespread across the pore, but RGWS power within localized regions can still increase in an individual spectral line. The later response is nevertheless dominated by an enhancement of the 4.2--7.0\,mHz oscillations across all eight lines.

The threshold-independent test confirms the overall late shift: total Fourier power increases in the 4.2--7.0\,mHz band in all eight lines and decreases in the 7.0--11.5\,mHz band in all eight lines. For Fe~{\sc ii}~328.1304\,nm, the upper-band Fourier result differs from the RGWS ratio. This difference reflects the distinct quantities measured: the Fourier calculation includes all band power, whereas the RGWS ratio includes only power that passes the wavelet-significance criterion.

\subsection{Velocity as a secondary check}
\label{subsec:velocity}

As a secondary test using a different observable, we repeated the RGWS analysis with the fitted line-center velocities. For this matched comparison, both intensity and velocity were reduced to the same $5\times5$-pixel spatial-block median time series before applying the same equal-duration intervals, frequency bands, and 95\% RGWS significance criterion. This matched block-level calculation is used only to compare the direction and statistical support of the intensity and velocity changes. The pixel-level intensity ratios shown in Figure~\ref{fig:band-power} remain the primary measurements.

In the late post-event interval, the 4.2--7.0\,mHz velocity power increased relative to its pre-event value in seven of the eight lines, compared with an increase in intensity power in all eight lines. Five of these velocity-power increases have 95\% bootstrap confidence intervals entirely above unity. Cu~{\sc i} and Fe~{\sc ii}~328.5421\,nm show increases whose confidence intervals include unity and are therefore not statistically distinguishable from no change. Fe~{\sc ii}~328.1304\,nm is the only exception, showing increased intensity power but reduced velocity power. As discussed in Section~\ref{subsec:fitting}, the distinct red-wing component seen near the flare peak is not evident as a separate component in the inspected early or late post-event profiles. Nevertheless, the fitted velocity of Fe~{\sc ii}~328.1304\,nm remains especially sensitive to changes among its absorption, central-reversal, and emission components.

The early post-event velocity response remains mixed, as does the corresponding intensity response. The velocity measurements therefore provide supporting evidence from a different observable for a delayed regional enhancement of the 4.2--7.0\,mHz oscillations, although the response is not identical in every line. The velocity measurements do not show a consistent reduction above 7\,mHz and therefore do not independently support the higher-frequency reduction found in intensity.

\section{Discussion and Conclusions}
\label{sec:discussion}

The compact C8.2 flare produced a clear but strongly line-dependent response across the selected near-ultraviolet spectrum. All eight fitted line cores brightened, but widespread full emission was confined to the three Fe~{\sc ii} lines. Tracking the evolving line centers was important for recovering the weaker responses, particularly in Cu~{\sc i}, because fixed-wavelength measurements can partly miss changes in shifted or evolving profiles.

The oscillatory response was not a uniform increase following the flare. During the early post-event interval, both the band-averaged detection fraction and intensity power of the 4.2--7.0\,mHz oscillations showed enhancement in some lines and suppression in others. A more consistent pattern developed later: the band-averaged detection fraction and intensity power in this band exceeded their pre-event values in all eight lines. Above 7\,mHz, the band-coverage fraction decreased in every line and intensity power decreased in seven, while Fe~{\sc ii}~328.1304\,nm maintained a localized enhancement. We therefore describe the result as a delayed, frequency-selective reorganization of the detected oscillatory power within the existing pore oscillatory state, rather than uniform broadband excitation. This interpretation neither requires the flare to have generated a new oscillation train nor implies that the observed oscillations triggered the flare.

Earlier spectral-line and ultraviolet imaging studies found flare-related excitation, changes in characteristic period and location, and delayed intensity or velocity responses in sunspots \citep{2007ApJ...670L.147K,2021MNRAS.503.2444M,2024MNRAS.527.5916M,2024ApJ...961..231W}. The present observations add simultaneous measurements in eight spectral lines, distinguish how commonly oscillations are detected from the magnitude of their oscillatory power, and resolve the evolution between two post-event intervals. The line-to-line differences are therefore part of the result: observations in only one line or one post-event interval could indicate either suppression or enhancement for this same event.

The response was not confined to the most strongly brightened pixels. Outside the strong-response mask, the late 4.2--7.0\,mHz band-coverage fraction increased in six of the eight lines, compared with all eight inside it. In the 7.0--11.5\,mHz band, the band-coverage fraction decreased throughout the selected pore region. The oscillatory reorganization therefore extended beyond the strongest flare brightening, although its lower-frequency component was more consistent among the lines inside that region. The flare ribbon crossed part of the selected pore region. This direct spatial overlap, together with the timing of the pore-wide oscillatory changes, supports a connection with the event but does not by itself establish causation. Three-minute oscillations can themselves show temporal modulation, including recurring trains with varying power and frequency \citep[e.g.,][]{2012A&A...539A..23S}. Such intrinsic variability may therefore contribute to the differences measured between our approximately 11-minute intervals. However, the transition from a mixed early response to a common late 4.2--7.0\,mHz enhancement across all eight lines, together with the independent Fourier and line-of-sight velocity results, shows that the measured change represents a robust reorganization of the oscillatory spectrum. While these observations are consistent with a flare-related modification, intrinsic temporal variability cannot be completely excluded for this single event.

Several related effects may contribute. Within the part of the pore crossed by the flare ribbon, flare heating can modify the sound speed, atmospheric scale heights, ionization balance, and radiative relaxation time, thereby changing the effective acoustic cutoff and the transmission of slow magnetoacoustic disturbances. Changes in magnetic-field inclination can also modify the cutoff for field-guided waves. Such magnetic changes could extend beyond the ribbon-brightened pixels and influence wave propagation across a wider part of the pore. Thermal or magnetic restructuring may shift reflecting layers or change the length and quality of an atmospheric resonant cavity \citep{2015LRSP...12....6K, 2020NatAs...4..220J, 2021NatAs...5....5J}. Within the ribbon-affected atmosphere, cooling plasma and post-flare downflows may provide an additional disturbance capable of exciting the natural chromospheric response \citep{2016ApJ...821L..30K,2024MNRAS.527.5916M}. These processes are not mutually exclusive, and the present observations do not determine which one dominates.

The selected lines sample different but overlapping parts of the perturbed atmosphere, and their measured responses depend on opacity, source function, ionization state, and sensitivity to temperature, density, and velocity perturbations. Lines with similar nominal formation heights are therefore not expected to show identical oscillatory spectra, particularly during a flare when their response functions can shift or broaden \citep{2019ApJ...870..114S,2026MNRAS.549g1001C}. Their ordering by flare enhancement should therefore not be interpreted as a strict ordering in formation height. Establishing the atmospheric response of each line will require radiative-hydrodynamic modeling and non-LTE spectral synthesis \citep{2016ApJ...832..147K, 2017ApJ...850...36C}.

The pre-event and late post-event states are likewise not expected to be identical. Figure~\ref{fig:flare} shows that most line-core intensities have returned close to their pre-event values by the early post-event interval, although Fe~{\sc ii}~327.7358 and Fe~{\sc ii}~328.1304\,nm retain modest enhancement. During the late post-event interval, these two lines remain slightly above their pre-event levels, whereas the other six, including Fe~{\sc ii}~328.5421\,nm, lie slightly below them. This indicates that the strongest radiative perturbation had subsided, although the small late offsets suggest that the line-forming atmosphere had not returned uniformly to its pre-event thermodynamic state. These offsets do not establish whether they result from post-flare evolution or a pre-event reference that was already slightly elevated, nor whether magnetic geometry had returned to its pre-event state. The late interval begins only about 15 minutes after the defined end of the flare interval, and each analysis interval contains approximately 11 minutes of data. The late measurement is therefore a finite sample of the post-event atmosphere rather than evidence of a permanent new equilibrium. The persistence of the oscillatory changes into this interval is consistent with a longer-lived change in magnetic inclination or waveguide geometry, although thermodynamic differences between the pre-event and late post-event atmospheres may also contribute. The strong pore power found from the longer observing sequence by \citet{2026ApJ..1005L..69J} is compatible with a short flare-related evolution embedded within longer-lasting pore oscillations.

The matched LOS-velocity analysis provides secondary support for the main intensity result. Late 4.2--7.0\,mHz velocity power increased in seven lines and was statistically clear in five. Fe~{\sc ii}~328.1304\,nm was the exception, but its fitted velocity is especially sensitive to its complex absorption, central-reversal, and emission components. The velocity measurements do not show the common higher-frequency reduction found in intensity, so that part of the response remains specific to the intensity measurements.

The present Stokes-$I$ observations cannot distinguish uniquely between thermodynamic changes and magnetic restructuring. A follow-up analysis will combine a larger selection of SUSI lines with the simultaneous spectropolarimetric measurements. Stokes inversions and time-dependent response functions will be required to test whether changes in field inclination, cutoff frequency, or reflecting layers account for the delayed response. Radiative-hydrodynamic modeling with non-LTE spectral synthesis would also help determine how flare heating, chromospheric condensation, and line-specific responses contribute to the observed differences.

In summary, over the tens of minutes following the line-core intensity peak of the compact C8.2 flare, the pore showed a line- and frequency band- dependent reorganization of its detected oscillatory power. Intensity power increased relative to the pre-event value in the 4.2--7.0\,mHz band in all eight lines. In the 7.0--11.5\,mHz band, intensity power decreased in seven lines, while RGWS detections became less widespread in all eight. These changes are consistent with a flare-related modification of the atmospheric waveguide or resonant response, while the relative contributions of thermodynamic and magnetic changes remain unresolved.

\begin{acknowledgments}
SJ and DBJ acknowledge support from the UK Science and Technology Facilities Council (STFC) through consolidated grants ST/T00021X/1 and ST/X000923/1 and for the small award UKRI5116. SJ also received support from the Rosseland Centre for Solar Physics (RoCS), University of Oslo, Norway. DBJ further acknowledges funding from the Leverhulme Trust (Research Project Grant RPG-2019-371) and from the UK Space Agency via the National Space Technology Programme (grant SSc-009). LF acknowledges support from UK Science and Technology Facilities Council (STFC) from consolidated grant ST/X000990/1. ROM would like to thank the European Office of Aerospace Research and Development (FA8655-22-1-7044-P00001) and the UK's Science \& Technology Facilities Council (ST/X000923/1) for supporting this research. PJAS acknowledges support from Conselho Nacional de Desenvolvimento Científico e Tecnológico (CNPq) grant 302740/2026-0. 
We wish to acknowledge scientific discussions with the Waves in the Lower Solar Atmosphere (WaLSA; \href{https://WaLSA.team}{www.WaLSA.team}) team, which has been supported by the Research Council of Norway (project no. 262622), The Royal Society (award no. Hooke18b/SCTM; \citealt{2021RSPTA.37900169J}), and the International Space Science Institute (ISSI Team 502).
\textsc{Sunrise iii} is supported by funding from the Max-Planck-F\"orderstiftung (Max Planck Foundation), NASA under Grants \#80NSSC18K0934 and \#80NSSC24M0024 (``Heliophysics Low Cost Access to Space'' program), and the ISAS/JAXA Small Mission-of-Opportunity program and JSPS KAKENHI Grant Numbers JP18H05234 and JP23K25916. This research has received financial support from the European Union's Horizon 2020 research and innovation programme under grant agreement No.~824135 (SOLARNET) and No.~101097844 (WINSUN) from the European Research Council (ERC). It has also been funded by the Deutsches Zentrum f\"ur Luft- und Raumfahrt e.V.\ (DLR, grant no.~50~OO~1608). The Spanish contributions have been funded by the Spanish MCIN/AEI under projects RTI2018-096886-B-C5 and PID2021-125325OB-C5, and from ``Center of Excellence Severo Ochoa'' awards to IAA-CSIC (SEV-2017-0709, CEX2021-001131-S), all co-funded by European REDEF funds, ``A way of making Europe''.
\end{acknowledgments}

\appendix
\section{Multichannel AIA and GOES context}
\label{app:aia-goes}

\begin{figure*}
    \centering
    \includegraphics[width=\textwidth]{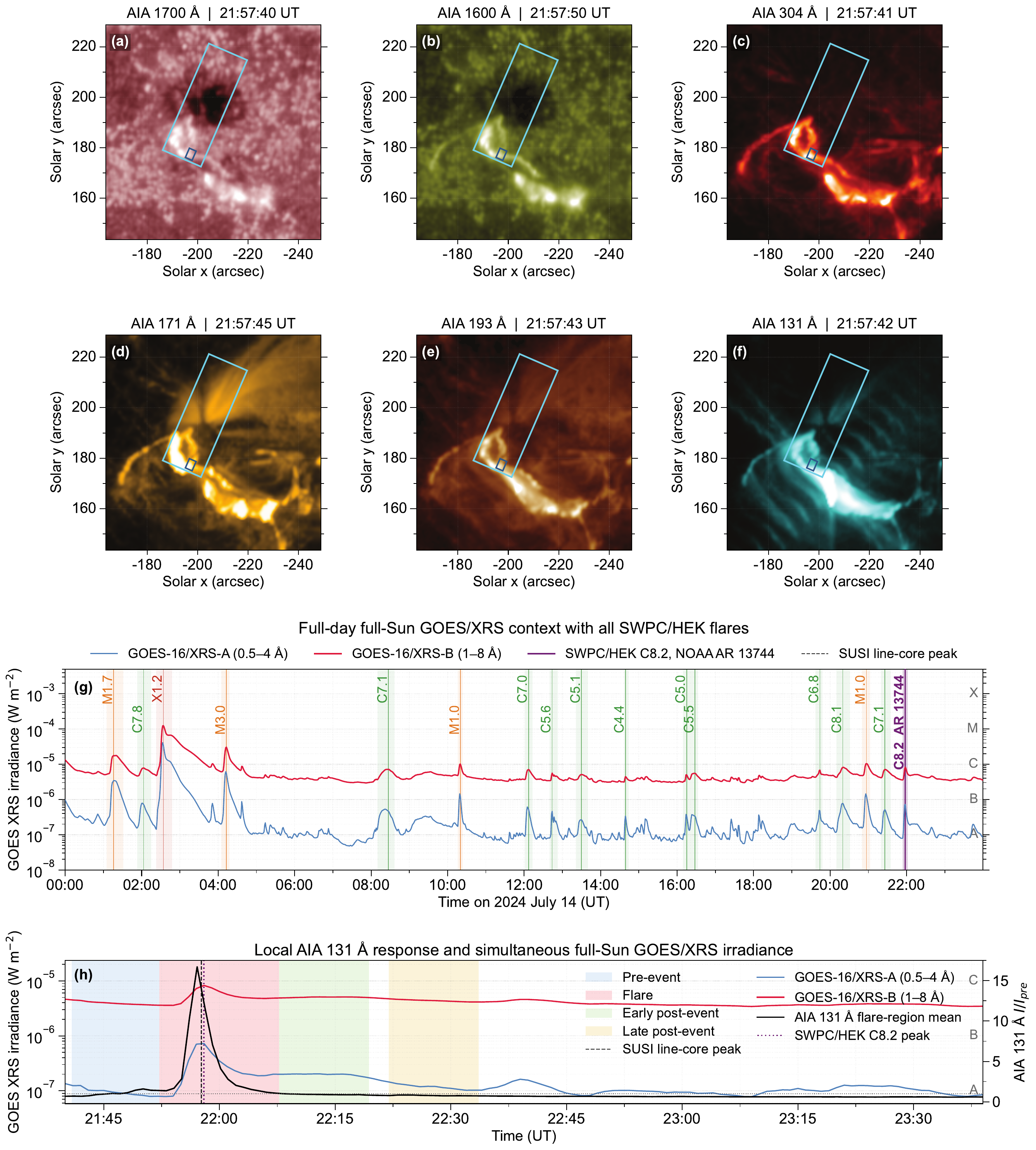}
    \caption{Multichannel imaging and full-Sun soft-X-ray context for the compact C8.2 flare. Panels (a)--(f) show co-registered SDO/AIA 1700, 1600, 304, 171, 193, and 131\,\AA\ images nearest to the SUSI line-core peak. The cyan and dark-blue outlines mark, respectively, the SUSI/SJI field of view and the analyzed pore region used in Figure~\ref{fig:context}. Each image has an independent asinh display scale. Panel (g) shows the full-day GOES-16/XRS-A and XRS-B irradiances to place the relatively small C8.2 enhancement in the context of the other solar activity recorded that day. The colored intervals, peak markers, and rotated labels identify all SWPC/HEK flare entries on 2024 July 14, irrespective of source active region; the C8.2 event assigned to NOAA active region 13744 is highlighted. The black dashed line marks the SUSI line-core peak, and the A--X reference levels apply to XRS-B. Panel (h) combines the local AIA 131\,\AA\ intensity, normalized by its pre-event median, with the simultaneous full-Sun XRS-A and XRS-B irradiances. The shaded regions mark the pre-event, flare-related, early post-event, and late post-event intervals defined in Section~\ref{subsec:intervals}. The black dashed and purple dotted lines mark the SUSI line-core and SWPC/HEK C8.2 peak times, respectively.}
    \label{fig:aia-goes-context}
\end{figure*}

We examined co-registered SDO/AIA images in six ultraviolet and extreme-ultraviolet passbands near the SUSI line-core peak (Figure~\ref{fig:aia-goes-context}). The same SUSI/SJI field of view and analyzed pore region used in Figure~\ref{fig:context} are marked in every image. Each passband is scaled independently for display, so the image brightness cannot be compared directly between channels. We also formed a 131\,\AA\ light curve from the compact flare region and divided it by its pre-event median.

For the soft-X-ray context, we used the science-quality, one-minute GOES-16/XRS Level-2 irradiances in XRS-A (0.5--4\,\AA) and XRS-B (1--8\,\AA). GOES/XRS measures the full-Sun irradiance and cannot be filtered by source active region. The full-day context is therefore useful for showing the background solar activity against which the relatively small C8.2 enhancement is observed, while the spatial association with the SUSI target must instead come from the SWPC/HEK source-region assignment and the co-spatial AIA response. We therefore show all 17 SWPC flare entries returned by HEK for 2024 July 14, irrespective of their reported source. The catalog contains a C8.2 event assigned to the observed NOAA active region 13744, starting at 21:54\,UTC, peaking at 21:58\,UTC, and ending at 22:02\,UTC. The catalog peak follows the SUSI line-core peak by approximately 20\,s. The detailed curves show a small contemporaneous increase in both GOES passbands, although the disk-integrated signal lies on the background of earlier activity. The C8.2 class refers to the absolute full-Sun XRS-B peak; the association with the SUSI target comes from the SWPC/HEK source-region assignment and the spatially coincident AIA response. The AIA images provide the local morphology of the event, but their intensities alone are not used as a temperature diagnostic.

\bibliography{article}{}
\bibliographystyle{aasjournalv7}

\end{document}